\documentclass[11pt]{article}
\def\be{\begin{equation}}
\def\ee{\end{equation}}
\def\ba{\begin{eqnarray}}
\def\ea{\end{eqnarray}}

\usepackage{epsfig}
\begin{document}
\begin{titlepage}
\vspace{4cm}
\begin{center}{\Large \bf Reply to \\ Comment on\\
''Quantum secret sharing based on a reusable Greenberger-Horne
-Zeilinger states as secure carriers''
}\\
\vspace{1cm} V. Karimipour\footnote{email: vahid@sharif.edu},\\
\vspace{1cm} Department of Physics, Sharif University of Technology,\\
P.O. Box 11365-9161,\\ Tehran, Iran
\end{center}
\vskip 3cm

\begin{abstract}
We show that the criticism of a recent comment \cite{ch2} on the
insecurity of a quantum secret sharing protocol proposed in
\cite{v2} is based on a misconception about the meaning of security
and hence is invalid. The same misconception also appears in another
comment of the authors \cite{ch1} on the security of an
entangled-based quantum key distribution protocol \cite{zhang,v1}.
\end{abstract}
\end{titlepage}
In a couple of comments \cite{ch2}, \cite{ch1}, it has been argued
that the protocols of quantum key distribution proposed in
\cite{zhang} and \cite{v1}, and the quantum secret sharing protocol
proposed in \cite{v2} are insecure. The basic idea of these comments
is based on a strategy in which Eve entangles herself with the state
already possessed by Alice and Bob and by suitable manipulations
intercepts the odd-numbered bits without being
recognized by Alice and Bob.\\
In the original articles \cite{v1,v2,zhang}, the possibility of
entanglement of Eve with the states already possessed by Alice and
Bob has been taken into account and methods for preventing her from
acquiring useful information has been devised. However the authors
of these two comments show that Eve can perform suitable operations
to intercept only the odd-numbered bits without being recognized by
Alice and Bob. Hence they conclude that the protocols are insecure.\\

While I agree with the mathematical analysis of these comments, I
should point out that this analysis by no means implies the
insecurity of the above mentioned protocols, for a very simple
reason: It is not a random sequence of bits which are intercepted by
Eve, but a fixed subset, known to the two legitimate parties. The
legitimate parties can simply discard the odd-numbered bits when
they want to establish their key. They can even use their protocol
as before for communicating {\it{predetermined}} messages, by
encoding their messages in the even round bits,
interspaced by stray random bits in the odd rounds.\\

To put this in the simplest form, they can communicate the word \be
R E P L Y\ee through the following sequence

\be X\ \  R\ \  X\ \ E\ \ X\ \ P\ \ X\ \ L\ \ X\ \ Y\ \, \ee in
which $X$ stands for stray bits whose interception by Eve does not
convey any information to her. \\

These protocols could have been proved insecure only if Eve could
intercept a sequence of bits unknown to the two parties.\\
The only viable result of the above comments is that they show that
the rate of information transmission (measured against the actual
number of bits transmitted) is not equal to $1$, as previously
thought in the above references, but is one half that value. This
value is now equal to the value in the BB84 protocol in which half
of the bits are discarded due to the mismatch of the directions of
spin measurements.

{}
\end{document}